\documentclass[fleqn,twoside]{article}
\usepackage{espcrc2}

\usepackage{pslatex}

\readRCS
$Id: ll08.tex,v 1.2 2004/02/24 11:22:11 spepping Exp $
\usepackage{graphicx}
\usepackage[figuresright]{rotating}

\newcommand{\AmS}{{\protect\the\textfont2
  A\kern-.1667em\lower.5ex\hbox{M}\kern-.125emS}}

\title{
\vspace*{-26mm} \rightline{ {\normalsize{DESY 08--102}}}
\vspace*{-2mm} \rightline{ {\normalsize{TTP 08-31}}}
\vspace*{-2mm} \rightline{ {\normalsize{SFB/CPP-08-52}}}
\vspace*{-2mm} \rightline{ {\normalsize{July 2008}}}
\vspace*{+6mm}
Automating dipole subtraction\thanks{Presented by K.H. at 
{\it{Loops and Legs in Quantum Field Theory}}, 20--25 April 2008, Sondershausen (Germany).}}

\author{K. Hasegawa\address[MCSD]{Deutsches Elektronen-Synchrotron  
     DESY, Platanenallee 6, D-15738 Zeuthen, Germany},
        S. Moch\addressmark
        , and
        P. Uwer\address{Institut f\"ur Theoretische Teilchenphysik, 
              Universit\"at Karlsruhe, D-76128 Karlsruhe, Germany}}

\begin{document}

\begin{abstract}
We report on automating the Catani-Seymour dipole subtraction 
which is a general procedure to treat infrared divergences in real 
emission processes at next-to-leading order in QCD. 
The automatization rests on three essential steps: 
the creation of the dipole terms, the calculation of the color linked
squared Born matrix elements, 
and the evaluation of different helicity amplitudes. 
The routines have been tested for a number of complex processes, such as the
real emission process $gg \rightarrow t\bar{t}ggg$. 
\vspace{1pc}
\end{abstract}

\maketitle

\section{INTRODUCTION}
QCD as the gauge theory of the strong interaction allows to predict cross sections 
for hard scattering reactions which, at a hadron collider, typically involve 
high multiplicities of colored partons.
In the perturbative approach calculations based on exact QCD matrix elements 
at leading order (LO) provide first estimates 
for cross sections and differential distributions.
However, in the complicated environment of a hadron collider precision
calculations to next-to-leading-order (NLO) in QCD 
are often needed in order to reliably predict (and separate) the Standard Model
background from possible new physics signals.
In the era of LHC this has triggered a lot of activity concerning NLO QCD corrections to multi-particle reactions, 
see e.g. Refs.~\cite{Buttar:2006zd,Bern:2008ef}.

The salient feature of NLO corrections is the presence of virtual and real
emission contributions. Virtual-loop corrections exhibit both 
ultraviolet (UV) and soft and collinear divergences which we call IR 
divergencies in the following.   
The real corrections contain only IR divergencies from 
soft and collinear emissions.
Upon summation of the two parts all IR divergences 
cancel (for so-called IR safe observables)~\cite{Bloch:1937pw,Kinoshita:1962ur,Lee:1964is}.
Since virtual and real corrections have different phase space 
integrals, these cancellations are not always trivial. 
In the Catani-Seymour dipole formalism~\cite{Catani:1996vz,Phaf:2001gc,Catani:2002hc}, 
the IR divergences of virtual and real corrections are treated separately 
by subtracting suitable dipole terms so that each of the contributions becomes individually finite.
The dipole terms are constructed systematically relying on the
universal nature of soft and collinear limits in QCD. 
Thus the method allows for a general treatment of IR divergences to NLO in QCD.

Current applications in phenomenology consider processes with six or more
parton legs~\cite{Buttar:2006zd,Bern:2008ef} which require about one hundred dipole terms.
These calculations are rather tedious and since 
the algorithm underlying the Catani-Seymour dipole subtraction 
is a combinatorial one automatization is favored.
The construction of the complete subtraction terms relies to a large extent 
on squaring color correlated Born amplitudes and dressing them with the corresponding dipoles.
To that end, we can use existing software for the automatic evaluation of Born amplitudes~\cite{%
Stelzer:1994tk,Maltoni:2002qb,Hahn:2000kx,Mangano:2002ea,Krauss:2001iv,Boos:2004kh,Kanaki:2000ey,Papadopoulos:2000tt}
by means of a suitable interface. Thus, the completely automatic generation of all subtraction terms in the
dipole formalism becomes feasible.

In related work Ref.~\cite{Gleisberg:2007md} recently reported on details of an automatization, 
although the code is unpublished.
Ref.~\cite{Seymour:2008mu} made code publicly available. 
However, all process dependent information, i.e. the (color correlated) Born
squared matrix elements, still have to be provided by the user. 
Automating this step is precisely what we are aiming at in the present article.

\section{ALGORITHM}
\label{sec:algo}
In this section we briefly review the algorithm of the dipole subtraction 
with particular emphasis on the features of the real emission contributions
for a given process.

{\bf 1.} Choose all possible emitter pairs from the external legs.
In the dipole subtraction, the root of the splitting of the quarks and gluons 
is called emitter. 
For convenience we call the two fields into which an emitter splits, emitter pair. 
We use indices, $i, j,$ and $k$, for fields in a final state and,
respectively, indices, $a$ and $b$, for an initial state field. 
The quark (anti-quark) and gluon are denoted by $f$ ($\bar{f}$) and $g$.
In case both partons of an emitter pair are in the final state, 
possible combinations are $(i,j) = (1)(f,g),(2)(g,g),(3)(f,\bar{f})$.
In case of one parton in the initial and the other in the final state, 
we have $(a,i) = (4)(f,g),(5)(g,g),(6)(f,f),(7)(g,f)$. 
There are also the other combinations where the quarks are replaced by the anti-quarks
in the cases, (1), (4), (6), and (7). 

{\bf 2.} Choose all possible spectators for each emitter pair. 
The spectator is one external field which is different from both fields
of the emitter pair.
For a spectator in the final (initial) state denoted by $k$ ($b$), this condition means $k \not= i, j$ ($b \not= a$). 
It emerges from a special feature of the subtraction formalism namely 
that the color factors of the square terms $|\mbox{M}_{i}|^{2}$ are expressed
through the ones of the interference terms $\mbox{M}_{i}\mbox{M}_{j}^{\ast} (i \not= j)$ 
due to color conservation.

{\bf 3.} Construct the dipole terms from the chosen combinations of emitter and spectator. 
The previous steps provide all such combinations as pairings 
(emitter, spectator)$=(ij,k),(ij,b),(ai,k)$, and $(ai,b)$.
Each case corresponds to one dipole term, $\mbox{D}_{ij,k}, \mbox{D}_{ij}^{a}, \mbox{D}^{ai}_{k},$ and $\mbox{D}^{ai,b}$, 
respectively, and explicit expressions are given in~\cite{Catani:1996vz,Catani:2002hc}. 
For example, the dipole term $\mbox{D}_{ij,k}$ in the massless case reads,
\begin{eqnarray}
  \mbox{D}_{ij,k} = 
  \frac{-1}{2 p_i \cdot p_j} \langle ij,k |\frac{ \mbox{T}_k \cdot \mbox{T}_{ij}}
  {\mbox{T}_{ij}^2} \mbox{V}_{ij,k}| ij,k\rangle 
  \, .
\end{eqnarray}
The quantity, $\langle ij,k \ | \mbox{T}_k \cdot \mbox{T}_{ij} | \ ij,k\rangle $, is called 
color linked Born squared matrix element (CLBS). 
It is given by the Born amplitude squared with two additional color operator insertions
at the emitter and spectator legs. 
The color operator $\mbox{T}$ denotes either a fundamental $t^{a}_{ij}$ or an adjoint $f^{abc}$ generator, 
depending on the parton type (i.e. quark, anti-quark or gluon). 
The quantity $\mbox{V}_{ij,k}$ is the so-called dipole splitting function.
In case of the emitter being a quark, for example, $(ij,k)=(fg,k)$ the
splitting function is diagonal in the spin space of the quark: 
\begin{eqnarray}
  {\lefteqn{
      \langle s| \mbox{V}_{fg,k} |s'\rangle  = }} 
  \nonumber \\ 
  && 
  8\pi C_{F} \alpha_{s}
  \biggl[ \frac{2}{1-z_{i} (1-y_{ij,k})} 
  -(1+z_{i}) \biggr] \delta_{ss'}
  \, ,
\end{eqnarray}
where $z_i$ and $y_{ij,k}$ are functions of the external momenta as,
\begin{eqnarray}
  z_i &=& \frac{p_i \cdot p_k}{p_j \cdot p_k+p_i \cdot p_k} 
  \, ,
  \\
  y_{ij,k} &=& \frac{p_i \cdot p_j}{p_i \cdot p_j+p_j \cdot p_k+p_k \cdot p_i} 
  \, .
\end{eqnarray}
In case of the emitter being a gluon, for example, $(ij,k)=(f\bar{f},k)$, it
does exhibit a correlation with the gluon helicity according to
\begin{eqnarray}
  \label{splithel}
  {\lefteqn{
    \langle \mu|\mbox{V}_{f\bar{f},k}|\nu\rangle  = 8\pi T_R \alpha_{s} \biggl[ -g^{\mu \nu} 
  }}
  \nonumber 
  \\ 
  && 
  -\frac{2}{p_ip_j}(z_i p_i^{\mu} - z_j p_j^{\mu})(z_i p_i^{\nu} - 
  z_j p_j^{\nu}) \biggr]
  \, ,  
\end{eqnarray}
which has to be treated accordingly (see for example Ref.~\cite{Weinzierl:1998ki}).

In summary, these three steps generate all dipole terms 
to subtract all IR divergences of a real emission process at NLO. 
The construction of the integrated dipole terms is less involved 
and proceeds in complete analogy~\cite{Catani:1996vz,Catani:2002hc}, 
thus we skip a detailed discussion here.

\section{CODE STRUCTURE}
Here we briefly sketch our implementation of the dipole subtraction
formalism. There is a large freedom how such an implementation can be
done. In Ref.~\cite{Dittmaier:2007wz} for example the implementation
was done in form of two independent C/C++ libraries providing all the
necessary functions to evaluate the dipole terms. The slight
disadvantage of this approach is that the produced code is non-local
and that there is some redundancy in the calculation. In the present
work we follow a different approach. The main idea here is to have
code generator which will produce an optimized flat code which can be
further optimized by the compiler. 
To do so we haven chosen to interface a Mathematica program with 
MadGraph~\cite{Stelzer:1994tk,Maltoni:2002qb}.

\subsection{Mathematica code}
\label{sec:math}
We implement the creation of the dipole terms in Mathematica. 
With a given (real emission) parton scattering process as an input, 
the Mathematica code automatically writes down all dipole terms needed at NLO.
It provides all expressions explicitly except for the CLBS. 
The code creates the dipole terms in an order according to the kind of the
emitter pairs, i.e. the seven combinations of $(i,j)$ or $(a,i)$ listed in Sec.~\ref{sec:algo}.

The first group of dipole terms (dipole 1) 
are the ones with the emitter pairs, (1),(2),(4), and (5). 
These emitters reduce the NLO real emission process 
to a Born amplitude which is the LO contribution to a process with one less gluon in the final state. 
The second kind of dipole terms (dipole 2) has the emitter pair (3), while the 
third and fourth (dipole 3 and 4) have (6) and (7) as emitter pairs.

In order to demonstrate the code, let us discuss the example $g(a)g(b) \rightarrow u(1)\bar{u}(2)g(3)$. 
The code starts with the creation of the first dipole $\mbox{D}_{13,2}$ 
which belongs to the group dipole 1 and the output is written in the form,
\begin{eqnarray}
  \mbox{D}_{ijkfgk}(132) = \frac{-1}{2 p_i \cdot p_j} 
  \mbox{V}_{ijkfgk}(132)\frac{\mbox{B}1(132)}{\mbox{T}_{13}^2}
  \, ,
  \label{diout}
\end{eqnarray}
where $\mbox{V}_{ijkfgk}$ is the dipole splitting function and $\mbox{B}1$ denotes the CLBS. 
The indices, ${ijkfgk}(132)$, of $\mbox{D}$ and $\mbox{V}$ mean that 
(emitter, spectator)=($ij,k$)=(quark gluon, something)=(13,2). 
About the CLBS the code stores only the necessary information for the direct calculation. 
It writes each CLBS as $\mbox{B}\mbox{`i'}$ corresponding to $\mbox{Dipole} \  \mbox{`i'}$, 
where i can be 1,2,3, or 4. 
For instance, the output for $\mbox{B}1(132)$ in Eq.~(\ref{diout}) is returned in the form,
\begin{eqnarray}
  \mbox{B}1(132)= \mbox{B}1[ \{ \{\mbox{g},\mbox{pa} \}, \{\mbox{g},\mbox{pb} \} \}--> 
  \nonumber \\
  \vspace{4cm} \{ \{\mbox{u},\mbox{pijtil[1,3]} \}, \{\mbox{ubar},\mbox{pktil[2]} \} \}]
  \, ,
  \label{b1}
\end{eqnarray}
where $gg \rightarrow u\bar{u}$ is the reduced Born process. 
The function $\mbox{pijtil[1,3]}$ is the reduced momenta for the emitter and $\mbox{pktil[2]}$ for the spectator.
In general, for a given NLO real emission process with $n$ parton legs, each dipole term 
has a reduced Born squared matrix element with $(n-1)$ parton legs. 
The reduced $(n-1)$ external momenta are functions of the original $n$ external momenta. 
For example, the reduced momentum for an emitter in the dipole term
$\mbox{D}_{ij,k}$ reads
\begin{eqnarray}
  {\widetilde p}_{ij}^{\mu} = p_{i}^{\mu} + p_{j}^{\mu} - \frac{y_{ij,k} }{1-y_{ij,k}} p_{k}^{\mu}
  \, .
\end{eqnarray}
The Mathematica code provides explicit expressions for the reduced momenta of each
dipole in the output according to Eq.~(\ref{b1}). 

At the end of the run, the total number of generated dipoles is shown.  
The final output for the complete subtraction term 
is given as a C- or Fortran routine for numerical evaluation.
We have tested the generation of dipoles with our Mathematica code for  complex processes, 
obtaining, e.g. twenty seven dipole terms for the process $gg \rightarrow
u\bar{u}g$, eighty dipoles for $gg \rightarrow u\bar{u}d\bar{d}g$, 
and one hundred dipoles for $gg \rightarrow u\bar{u}ggg$.

\subsection{Interface to MadGraph}
\label{sec:mad}
As we see, the remaining ingredient at this stage is the CLBS appearing in all dipole terms. 
In case the emitter is a gluon, they also include the different helicity components of the CLBS 
for the emitter.
In order to obtain these quantities in an automatic way, it is advantageous to
use a publicly available software 
for automated LO calculations. 
We choose MadGraph for this purpose and interface our Mathematica program with 
the stand-alone version~\cite{Stelzer:1994tk,Maltoni:2002qb}. 

Let us briefly explain our interface to MadGraph to obtain the CLBS.  
In MadGraph, the color factors are separated from each diagram. In the
evaluation everything is expressed in terms of generators of the
fundamental representation.
A typical example is that the factor $f^{abc}$ of the gluon three point vertex
is rewritten in terms of the fundamental generator $t^{a}_{ij}$ due to the identity,
\begin{eqnarray}
  f^{abc} = -2 i \bigl( \ \mbox{Tr}[t^{a}t^{b}t^{c}] - \mbox{Tr}[t^{c}t^{b}t^{a}] \ \bigr) 
  \, .
\end{eqnarray}
The color factors of each diagram are sorted in an unique order and they are expressed in a sum of some terms. 
When a specific term of a diagram is identical to one of the other diagrams, it is combined as 
\begin{eqnarray}
  \mbox{M} = \sum_{a} C_{a} \mbox{J}_{a}
  \, ,
\end{eqnarray}
where $C_{a}$ denotes the independent color factors. 
Each $C_{a}$ has fundamental and adjoint color indices corresponding to the external quarks (anti-quarks) and the gluons, respectively.
$\mbox{J}_{a}$ is the joint amplitude, e.g. $\mbox{J}_{1}=+A_{1}-A_{3}+ \cdots$ where $A_{i}$ 
is the partial amplitude of $i$-th diagram (with the color factor stripped off).
The invariant matrix element squared is finally expressed in the form,
\begin{eqnarray}
  \label{eq:mesquared}
  |\mbox{M}|^{2} = \bigl(\vec{\mbox{J}}\bigr)^{\dagger} \ \mbox{CF} \ \vec{\mbox{J}}
  \, ,
\end{eqnarray}
where the color matrix $\mbox{CF}$ is defined as 
\begin{equation}
(\mbox{CF})_{ab}=\sum_{\mbox{\scriptsize color}}C_{a}^{\ast} C_{b}.   
\end{equation}

For the CLBS we need to evaluate Eq.~(\ref{eq:mesquared}) with an insertion of
two additional color operators to the emitter and spectator legs. 
This is precisely what our interface to MadGraph does.

Let us return to the example of $\mbox{B}1(123)$ from the previous subsection.
The reduced Born process $g(a)g(b) \rightarrow u(1)\bar{u}(2)$ has three diagrams and the
color factors are combined into two independent ones,
$(C_{1}, C_{2}) = ((t^{a} t^{b})_{12},(t^{b} t^{a})_{12})$. 
The components of the color matrix are written in the traces, 
$(\mbox{CF})_{11}=(\mbox{CF})_{22}=\mbox{Tr}[t^{b}t^{a}t^{a}t^{b}]$ and
$(\mbox{CF})_{12}=(\mbox{CF})_{21}=\mbox{Tr}[t^{b}t^{a}t^{b}t^{a}]$.
Then the color matrix is calculated as 
\begin{eqnarray}
  \mbox{CF} = \left( \begin{array}{rr}
      16/3 & -2/3  \\
      -2/3 & 16/3  \\
    \end{array}\right)
  \, .
\end{eqnarray}
For the CLBS $\mbox{B}1(123)$ we need the fundamental operator insertions into the legs 1 and 2.
The components of the color matrix $\mbox{CF}$ are modified to
$(\mbox{CF}')_{11}=\mbox{Tr}[t^{b}t^{a}t^{c}t^{a}t^{b}t^{c}]$ and
$(\mbox{CF}')_{12}=\mbox{Tr}[t^{b}t^{a}t^{c}t^{b}t^{a}t^{c}]$.
Then the modified color matrix is calculated as 
\begin{eqnarray}
  \mbox{CF}' = \left( \begin{array}{rr}
      1/9 & 10/9  \\
      10/9 & 1/9  \\
    \end{array}\right).
\end{eqnarray}
The subroutines of MadGraph for the color factor calculations are well structured and the
original routines to add the color factors $t^{a}_{ij}$ and $f^{abc}$ can be applied to the
additional color insertions for CLBS. 
We have realized the two color insertions in an automatic way and checked that MadGraph 
with our interface works for rather involved processes.
One of the most complex checks consists of the two color insertions into the process 
$g(a)g(b) \rightarrow u(1)\bar{u}(2)g(3)g(4)$. 
In MadGraph the normal color matrix for the process is a 24 by 24 matrix. 
Here we show only the first 15 components in the first row as
\begin{eqnarray}
{\lefteqn{
  \mbox{CF} = \frac{1}{54}(512, 8, -64,80, 8, -10,
}} 
  \\
  \nonumber 
  && 
   -1, -64, -64, 8, -1, -10, -1, 62, -10 , \cdots ).
\end{eqnarray}
Next, we perform two adjoint operator insertions into the legs 3 and 4, and the 
extended routines calculate the modified color matrix as
\begin{eqnarray}
    \label{cfmod}
{\lefteqn{
  \mbox{CF}' = \frac{1}{4}(8,0,8,16,0,-2,
}}
  \\
  \nonumber 
  && 
  0,8,-1,-1,1,2,-8,-7,1, \cdots )\, .
\end{eqnarray}
We have checked that the result in Eq.~(\ref{cfmod}) and the sum of all components agree 
with results of our independent private code. 

Next we briefly comment on the different helicity components of the CLBS for the gluon emitter. 
In MadGraph the gluon polarization vector is calculated by the subroutine `VXXXXX' of the 
HELAS library where the polarization vector is taken in the circular
polarization representation~\cite{Murayama:1992gi,Hagiwara:1990dw}. 
Then, it is favorable to calculate the dipole terms with the helicity correlation 
in the circular polarization basis. 
For example, in the previous process $g(a)g(b) \rightarrow u(1)\bar{u}(2)g(3)$, we take one dipole $\mbox{D}^{a3}_{1}$
and calculate it in the circular polarization basis as $\mbox{V}^{a3}_{1}(\lambda,\lambda') 
\mbox{B}1(g(\lambda,\lambda') g \rightarrow u\bar{u})$ where $\mbox{V}^{a3}_{1}(\lambda,\lambda')$ 
is constructed from the dipole splitting function in the basis of the Lorentz
indices as in Eq.~(\ref{splithel}) by multiplying the circular polarization vectors. 
The arguments $(\lambda,\lambda')$ are the ones for the different helicities of
the original and the complex conjugated amplitude for the emitter gluon. 
We have  completed the interface to obtain the CLBS with the different helicities 
as $\mbox{Bi}(\lambda,\lambda')$ and checked it in  
processes as the above $\mbox{B}1(g(\lambda,\lambda') g \rightarrow u\bar{u})$.

\subsection{Complete structure}
\begin{figure*}[htb]
\begin{center}
\includegraphics[width=11cm]{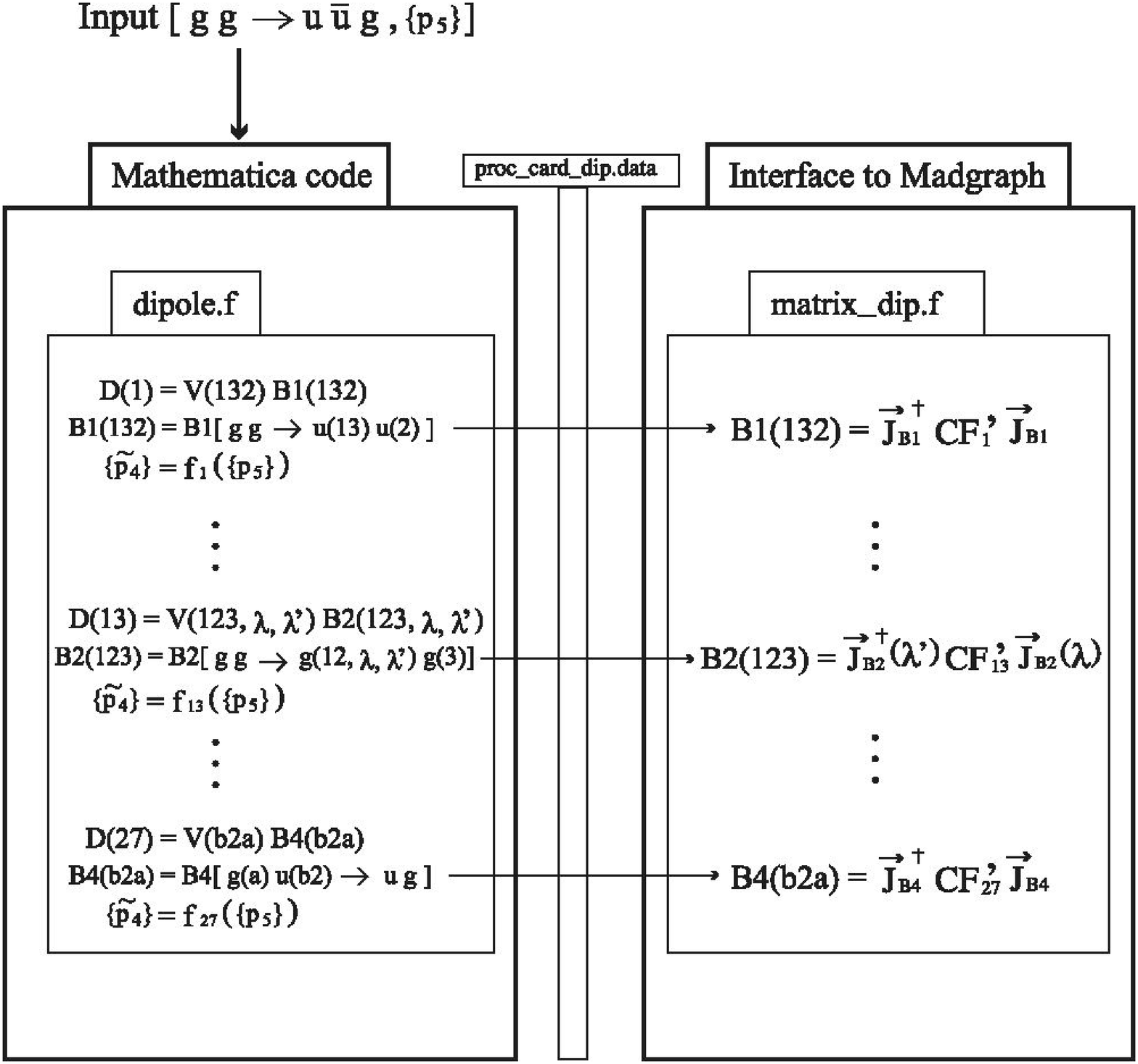}
\end{center} 
\caption{ \small 
  \label{fig1}}
The complete structure of the code is shown. The Mathematica generates 
all subtraction terms in the dipole formalism and returns output in C or
Fortran for numerical evaluation. The calculation of the CLBS is delegated to
MadGraph via an interface.
\end{figure*}
Let us finally wrap up by displaying the complete structure of our code 
shown in the flowchart in Fig.~\ref{fig1}. 
An user specifies a $n$-parton real emission process as an input to the Mathematica code 
as well as a set of the external momenta $\{p_{n}\}$.
The Mathematica program then generates all dipole terms in the appropriate order 
(see Sec.~\ref{sec:math}) along with the CLBS $\mbox{B}1, \ \cdots ,\mbox{B}4$. 
The created dipole terms are written as C- or Fortran routines to a file `dipole.c' or `dipole.f'.
In this file the explicit expressions for the dipole splitting function $\mbox{V}$ are contained. 
Together with each dipole, the information on its reduced kinematics as a
function of the external momenta is stored (see Sec.~\ref{sec:math}).

The necessary information to calculate each CLBS is transferred to MadGraph 
through the file `proc\_card\_dip.data'.
This data file is an imitation of the input file `proc\_card.data' of the original MadGraph. 
Our interface reads `proc\_card\_dip.data' and gets MadGraph 
to write Fortran routines for the evaluation of each CLBS in a file `matrix\_dip.f'.
As we see in Sec.~\ref{sec:mad}, the color matrix $\mbox{CF}$ 
is modified to $\mbox{CF}'$ due to the additional color insertions. 

Finally, the C- and/or Fortran codes in the two files can be used for the
numerical evaluation of all dipole terms as functions of the external momenta $\{p_{n}\}$. 
The sum of the dipole subtractions from the invariant matrix element squared
of the NLO real emission process reads,
\begin{eqnarray}
  \label{eq:finsub}
  |\mbox{M}(2 \rightarrow (n-2)\mbox{partons})|^{2} - \sum_{i} \mbox{D}(i)
  \, .
\end{eqnarray}
The invariant matrix element squared can be calculated with the original MadGraph
version and $i$ runs over all dipole terms. 
Eq.~(\ref{eq:finsub}) is finite upon integration over the phase space of the unresolved parton.
Thus, it can finally be integrated over the phase space by using standard Monte Carlo
techniques to obtain its contribution to an IR safe cross section.

\section{OUTLOOK}
We have reported on ongoing work to automate the Catani-Seymour dipole
formalism in order to calculate the subtracted invariant matrix element
squared Eq.~(\ref{eq:finsub}) in an automatic way.

The automatization essentially requires three ingredients: 
the automatic generation of all dipole terms, the calculation of the CLBS, and the evaluation of different helicity amplitudes.
The implementation of each of these tasks either in our Mathematica program or
in an interface to MadGraph has  been completed and the respective
routines have undergone sufficient checks.
We are now finalizing the user interface and the output format to achieve full automatization. 
At the same time we are checking our code for various massless real emission processes,
like $gg \rightarrow u\bar{u}g$, $gg \rightarrow u\bar{u}gg$, and $gg \rightarrow u\bar{u}ggg$,
as well as for massive processes, like 
$gg \rightarrow t\bar{t}g$, $gg \rightarrow t\bar{t}gg$, and $gg \rightarrow t\bar{t}ggg$ 
to obtain finite results for $|\mbox{M}|^{2} - \sum \mbox{D}$ in all soft and
collinear limits (and IR safe contributions to cross sections).
Once the reliability of our software has been fully established and
the code has been optimized for speed, it will be made publicly available.


\begin{thebibliography}{10}

\bibitem{Buttar:2006zd}
C.~Buttar {\it et al.},
\newblock hep-ph/0604120.
\newblock 

\bibitem{Bern:2008ef}
Z.~Bern {\it et al.}  [NLO Multileg Working Group],
\newblock arXiv:0803.0494 [hep-ph].
\newblock 

\bibitem{Bloch:1937pw}
F. Bloch and A. Nordsieck,
\newblock Phys. Rev. 52 (1937) 54.
\newblock 

\bibitem{Kinoshita:1962ur}
T. Kinoshita,
\newblock J. Math. Phys. 3 (1962) 650.
\newblock 

\bibitem{Lee:1964is}
T.D. Lee and M. Nauenberg,
\newblock Phys. Rev. 133 (1964) B1549.
\newblock 

\bibitem{Catani:1996vz}
S. Catani and M.H. Seymour,
\newblock Nucl. Phys. B485 (1997) 291, hep-ph/9605323.
\newblock 

\bibitem{Phaf:2001gc}
L. Phaf and S. Weinzierl,
\newblock JHEP 04 (2001) 006, hep-ph/0102207.
\newblock 

\bibitem{Catani:2002hc}
S. Catani, S. Dittmaier, M.H. Seymour, and Z. Trocsanyi,
\newblock Nucl. Phys. B627 (2002) 189, hep-ph/0201036.
\newblock 

\bibitem{Stelzer:1994tk}
T.~Stelzer and W.~F.~Long,
\newblock Comput.\ Phys.\ Commun.\  {\bf 81} (1994) 357, hep-ph/9401258.
\newblock 

\bibitem{Maltoni:2002qb}
F. Maltoni and T. Stelzer,
\newblock JHEP 02 (2003) 027, hep-ph/0208156.
\newblock 

\bibitem{Hahn:2000kx}
T. Hahn,
\newblock Comput. Phys. Commun. 140 (2001) 418, hep-ph/0012260.
\newblock 

\bibitem{Mangano:2002ea}
M.L. Mangano et~al.,
\newblock JHEP 07 (2003) 001, hep-ph/0206293.
\newblock 

\bibitem{Krauss:2001iv}
F. Krauss, R. Kuhn and G. Soff,
\newblock JHEP 02 (2002) 044, hep-ph/0109036.
\newblock 

\bibitem{Boos:2004kh}
E.~Boos {\it et al.}  [CompHEP Collaboration],
\newblock Nucl.\ Instrum.\ Meth.\  A {\bf 534} (2004) 250, hep-ph/0403113.
\newblock 

\bibitem{Kanaki:2000ey}
A. Kanaki and C.G. Papadopoulos,
\newblock Comput. Phys. Commun. 132 (2000) 306, hep-ph/0002082.
\newblock 

\bibitem{Papadopoulos:2000tt}
C.G. Papadopoulos,
\newblock Comput. Phys. Commun. 137 (2001) 247, hep-ph/0007335.
\newblock 

\bibitem{Gleisberg:2007md}
T. Gleisberg and F. Krauss,
\newblock Eur. Phys. J. C53 (2008) 501, arXiv:0709.2881 [hep-ph].
\newblock 

\bibitem{Seymour:2008mu}
M.H. Seymour and C. Tevlin,
\newblock (2008), arXiv:0803.2231 [hep-ph]
\newblock 

\bibitem{Weinzierl:1998ki}
S.~Weinzierl,
\newblock SACLAY-SPH-T-98-083.

\bibitem{Dittmaier:2007wz}
S.~Dittmaier, P.~Uwer and S.~Weinzierl,
\newblock Phys.\ Rev.\ Lett.\  {\bf 98}, 262002 (2007), hep-ph/0703120.
\newblock 

\bibitem{Murayama:1992gi}
H. Murayama, I. Watanabe and K. Hagiwara,
\newblock KEK-91-11.

\bibitem{Hagiwara:1990dw}
K. Hagiwara, H. Murayama and I. Watanabe,
\newblock Nucl.\ Phys.\  B {\bf 367} (1991) 257.
\newblock 

\end{thebibliography}
\end{document}